\documentstyle[twoside,12pt,epsf]{article}
\normalsize
\newcommand{\bdi}{\begin{displaymath}}
\newcommand{\edi}{\end{displaymath}}
\newcommand{\bfi}{\begin{figure}}
\newcommand{\efi}{\end{figure}}

\newcommand{\beq}{\begin{equation}}
\newcommand{\eeq}{\end{equation}}
\newcommand{\beqa}{\begin{eqnarray}}
\newcommand{\eeqa}{\end{eqnarray}}

\def\longbar#1{\setbox1=\hbox{$#1$}
\setbox2=\vbox{\hrule width 0.8\wd1}
\raise0.5\ht1\hbox{${\lower\dp1\box2}\atop\box1$}}  

\begin{document}

\begin{titlepage}

\begin{flushright}
\today
\end{flushright}

\vspace{1cm}
\begin{center}
{\Large \bf Zero modes in finite range magnetic fields}\\[1cm]
C. Adam* \\
Institut f\"ur theoretische Physik, Universit\"at Karlsruhe, 76128 Karlsruhe\\

\medskip

\medskip

B. Muratori**,\, C. Nash*** \\
Department of Mathematical Physics, National University of Ireland, Maynooth
\vfill
{\bf Abstract} \\
\end{center}

We find a class of Fermion zero modes of  Abelian Dirac operators in three
dimensional Euclidean space 
where the gauge potentials and the related magnetic fields are
nonzero only in a finite space region. 

\vfill

$^*)${\footnotesize  
email address: adam@particle.uni-karlsruhe.de, adam@pap.univie.ac.at} 

$^{**})${\footnotesize
email address: bmurator@fermi1.thphys.may.ie} 

$^{***})${\footnotesize
email address: cnash@stokes2.thphys.may.ie} 
\end{titlepage}

\section{Introduction}

Fermion zero modes of the Abelian Dirac operator in three dimensional
Euclidean space (i.e., of the Pauli operator) are a rather young subject of
study, and still many features remain unknown. The first example of such
a zero mode of the Pauli operator has been given in \cite{LoYa}, where
it was used to prove that one-electron atoms with sufficiently high
nuclear charge in an external magnetic field are unstable, see
\cite{FLL}. Further examples of zero modes were discussed in
\cite{AMN1}, where their relevance for QED was briefly mentioned
(this point is discussed in more detail in  \cite{Fry1,Fry2}).
In \cite{AMN2,AMN3} it was proven by explicit construction of a class of
examples that the phenomenon of zero mode degeneracy (i.e., Pauli
operators with more than one zero mode) occurs, and a relation between
the number of zero modes of a Pauli operator and the Hopf index of the
corresponding magnetic field was established. This point was further
elaborated in \cite{ES1}. 

In \cite{Elt1} an example of a zero mode was given where the
corresponding gauge potential (and magnetic field) are non-zero only
within a finite region of space (within a ball with finite radius).
In fact, this example belongs to the types of zero modes that were
discussed in \cite{AMN1}. Here we want to construct a whole class of
zero modes of Pauli operators where the related gauge potentials and
magnetic fields vanish outside a finite region of space. This
demonstrates that the possibility of having zero modes in magnetic
fields of finite range is not just a curiosity that is related to
some very special examples, but a rather generical feature
of the Pauli operator. We think that this observation is interesting
from a physical perspective as well, because
magnetic fields with finite range are precisely the types of
magnetic fields that may be realised experimentally. 

\section{Construction of the zero modes}

We want to study specific solutions of the equation
\beq
 -i\sigma_i \partial_i \Psi (x) 
=A_i (x) \sigma_i \Psi (x)
\eeq
(here $x=(x_1,x_2,x_3)$, $r=|x|$, 
and $\sigma_i$ are the Pauli matrices)
where $\Psi$ is square-integrable, $\Psi (x) \in L^2 ({\rm\bf R}^3)$,
 $A_i$ and $B_i =\epsilon_{ijk}\partial_j A_k$
are non-singular everywhere in ${\rm\bf R}^3$ and are different
from zero only in a finite region of space. Further $\Psi$, $A_i$ and
$B_i$ have to be smooth everywhere.
For this purpose, let us first observe that the spinor ($x_\pm \equiv x_1
\pm ix_2$)
\beq
\Psi^0 (x) =\frac{i}{r^3}\left( \begin{array}{c} x_3  
\\ x_+  \end{array} \right)
\eeq
solves the free Dirac equation
\beq
-i\vec\sigma\vec\partial \Psi^0 =0
\eeq
(the $i$ in (2) is chosen for later convenience). The spinor (2) is
singular at $r=0$ but it is well behaved for large $r$. So we might ask
whether there exist spinors that are equal to $\Psi^0$ outside a ball of
radius $r=R$ (where they solve the free Dirac equation, i.e., $A_i =0$
for $r>R$), whereas they differ from $\Psi^0$ inside $r=R$. Inside the ball
they are supposed to solve the Dirac equation for some nonzero $A_i$ such
that they are nonsingular and smooth everywhere. We shall find a whole
class of such zero modes among the zero modes that were discussed in
\cite{AMN1}, therefore we want to review the results of \cite{AMN1}
briefly. There the ansatz 
 \beq
\Psi =g(r)\exp (if(r)\frac{\vec x}{r} \vec\sigma )
\left( \begin{array}{c} 1  
\\ 0 \end{array} \right)
= g(r)[\cos f(r){\bf 1}+
i\sin f(r) \frac{\vec x}{r}  \vec\sigma ]
\left( \begin{array}{c} 1 
\\ 0  \end{array} \right)
\eeq
for the spinor lead to a zero mode for the gauge field
\beq
A_i =h(r) \frac{\Psi^\dagger \sigma_i \Psi}{\Psi^\dagger \Psi} 
\eeq
provided that $g(r)$ and $h(r)$ are given in terms of the independent
function $f(r)$ as ($' \equiv d/dr$)
 \beq
g' =-\frac{2}{r}\frac{t^2}{1+t^2}g.
\eeq
\beq
h=(1+t^2)^{-1}(t' +\frac{2}{r}t)
\eeq
where
\beq
t(r):= \tan f(r).
\eeq
A sufficient condition on $t(r)$ leading to
smooth, non-singular and $L^2$ spinors and
smooth, non-singular gauge potentials with finite energy ($\int (\vec
B)^2$) and finite Chern--Simons action ($\int \vec A \vec B$) is
\beq 
t(0)=0 \, , t(r) \sim c_1 r + o(r^2) \, \,  {\rm for} \, \, r \to 0
\eeq
\beq
t(\infty) =\infty
\eeq
which we shall assume in the sequel. Observe that in the limit $t\to
\infty$ $\vec A$ vanishes whereas $\Psi$ becomes $\Psi^0$. Therefore, if
we find some $t$ that become infinite at some finite $r=R$ in a smooth
way and stay infinite for $r>R$, we have found precisely what we
want. To get a more manageable condition, let us re-express things in
terms of
\beq
c(r):=\cos f(r)=\frac{1}{(1+t(r)^2)^{1/2}}
\eeq
which leads to
\beq
g' = -\frac{2}{r}(1-c^2)g
\eeq
\beq
h=c(-\frac{c'}{(1-c^2)^{1/2}} + \frac{2}{r}(1-c^2)^{1/2}).
\eeq
Further $c$ has to behave like
\beq
c(r) \sim 1-c_2 r^2 +\ldots \quad {\rm for} \quad r\to 0.
\eeq
Now let us assume that $c$ approaches zero in a smooth way for $r=R$ and
stays zero for $r\ge R$, and further $c'(r=R)=0$, $|c'' (r=R)|<\infty$. 
This implies that $\vec A=0$ for $r>R$ and that
\beq
g(r) =kr^{-2} \, ,\quad k=\exp (-2\int_0^R dr \frac{1-c(r)^2}{r})
\quad {\rm for}\quad r\ge R
\eeq
which precisely leads to $\Psi = {\rm const}\cdot \Psi^0$ for $r>R$, see
(4). 

Finally, let us give some examples, where we choose $R=1$ for convenience.
A first example is
\beq 
c(r) = (1-r^2)^2 \quad {\rm for}\quad r< 1 \, , \quad 
c(r) =0 \quad {\rm for}\quad  r\ge 1
\eeq
leading to
\bdi
g(r) = \exp (-4r^2+3r^4 -\frac{4}{3}r^6 +\frac{1}{4}r^8)  
\quad {\rm for}\quad r< 1 
\edi
\beq
g(r) = \exp (-\frac{25}{12}) \, r^{-2} \quad  {\rm for} \quad  r\ge 1 
\eeq
and
\bdi
h(r)= \frac{2 (1-r^2)^2(2 -4 r^2 + 4r^4 -r^6)}{(4 -6r^2 +4r^4 -r^6)^{1/2}} 
\quad {\rm for}\quad r< 1 
\edi
\beq
\quad h(r) =0 \quad {\rm for}\quad  r\ge 1.
\eeq
Another example is
\beq
c(r)= \exp (\frac{r^2}{r^2 -1}) \quad {\rm for}\quad r< 1 \, ,\quad
c(r) =0 \quad {\rm for}\quad r\ge 1.
\eeq
There is one difference between example (16) and example (19). Both lead
to $L^2$ zero modes (i.e., bound states), and both lead to magnetic
fields with are smooth, non-singular and have finite energy. However,
higher derivatives of $c$ in (16) are discontinuous, whereas all
derivatives of $c$ in (19) are smooth. In some situations (or for
mathematical reasons) it may be preferable to have only such $c$ that
have only smooth higher derivatives, then functions $c$ like in (16) may
be treated as follows with the help of functions like (19). 
Define a function $c_a$
\beq
c_a (r)= (1-r^2)^2 \exp (\frac{a r^2}{r^2 -1}) 
\quad {\rm for}\quad r< 1 \, ,\quad
c(r) =0 \quad {\rm for}\quad r\ge 1
\eeq
where $a\ge 0$.
For $a\ne 0$ $c_a$ is a $C^\infty$ function. 
In the limit $a\to 0$ $c_a$ is equal to the $c$ of (16).
Further, for $a$
sufficiently small, $c_a$ approximates the $c$ of (16) with arbitrary
precision. Therefore, the function $c_a$ with a sufficiently small $a$ 
may be used as a $C^\infty$ substitute for (16).

\section{Summary}

As should be clear from the above discussion, 
there is an infinite number of possible functions $c(r)$,
therefore already for the special ansatz (4) there exists a whole class
of zero modes in finite range magnetic fields.
One obvious generalisation of the above result is the existence of zero
modes in magnetic fields that are non-zero inside a ball $r<R_1$, zero
between $R_1$ and $R_2 > R_1$, non-zero again between $R_2$ and $R_3 >
R_2$, etc., forming an onion-like structure.

We started from the spherically symmetric ansatz (4), therefore the
finite regions where $\vec B\ne 0$ are all spherically symmetric balls. It is
plausible to assume that by relaxing the symmetry condition on the zero
modes one could find zero modes for magnetic fields which vanish outside
 finite regions of different shapes.

\end{document}